\documentclass[pra,twocolumn,aps,showpacs,nofootinbib,floatfix,groupedaddress]{revtex4-1}
\usepackage{amsmath}
\usepackage{amssymb}
\usepackage{graphicx}
\usepackage{verbatim}
\usepackage[colorlinks=true,linkcolor=blue,citecolor=blue,urlcolor=blue]{hyperref}
\usepackage{txfonts}
\usepackage{dsfont}
\usepackage{xr}

\begin{document}

\title{Optical control of the current-voltage relation in stacked superconductors}
\author{Frank Schlawin, Anastasia S. D. Dietrich and Dieter Jaksch}
\address{Clarendon Laboratory, University of Oxford, Parks Road, Oxford OX1 3PU, United Kingdom}
\email{frank.schlawin@physics.ox.ac.uk}

\begin{abstract}

We simulate the current-voltage relation of short layered superconductors, which we model as stacks of capacitively coupled Josephson junctions. 
The system is driven by external laser fields, in order to optically control the voltage drop across the junction.
We identify parameter regimes in which supercurrents can be stabilised against thermally induced phase slips, thus reducing the effective voltage across the superconductor. Furthermore, single driven Josephson junctions are known to exhibit phase-locked states, where the superconducting phase is locked to the driving field. 
We numerically observe their persistence in the presence of thermal fluctuations and capacitive coupling between adjacent Josephson junctions. 
Our results indicate how macroscopic material properties can be manipulated by exploiting the large optical nonlinearities of Josephson plasmons. 

\end{abstract}

\maketitle

\section{Introduction}

Recent years have seen tremendous progress in the optical control of solid state systems due to the advent of strong terahertz sources \cite{Kampfrath11, Kampfrath13, Nicoletti16}. Notable examples include the nonlinear driving of specific phonon modes in strongly correlated materials, aiming to control the electronic dynamics~\cite{Forst11, Subedi14, Knap15, Sentef16, Millis17}, in order to create phases of matter which are suppressed in the equilibrium. Such driving can be used to induce metal-insulator transitions~\cite{Rini07, Liu12}, excite synthetic magnetic fields \cite{Nova17}, and melt striped phases~\cite{Fausti11} or charge density waves~\cite{Forst14, Forst14b, Mankowsky17}. 
It can even give rise to transient superconducting-like phases \cite{Hu14, Mankowsky14, Mitrano16}.
All of these experiments triggered great interest in the emergent dynamics of driven many-body systems \cite{Bukov15}, which had previously been considered predominantly in the field of ultracold atoms~\cite{Bloch12}. They also raised the question to what extent electronic properties of materials can be controlled by strong laser pulses. 

In this endeavour, cuprate superconductors, and in particular the Josephson plasmon coupling between superconducting layers, have emerged as a particularly interesting platform. The Josephson coupling can be manipulated either by driving specific phonon modes to large amplitude, thus transiently distorting the lattice structure~\cite{Mankowsky14, Hoeppner15, Okamoto16, Okamoto17}, or by directly exploiting the large optical nonlinearity of the Josephson plasmon~\cite{Dienst11, Dienst13, Rajasekaran15, Rajasekaran18}. Recent theoretical proposals also investigated the possibility of laser cooling phase fluctuations in the system~\cite{Sam15}, and the optical control of currents in the system~\cite{Schlawin17}.

The full description of phase fluctuations and Josephson plasma waves in these systems requires solving nonlinear coupled sine-Gordon equations~\cite{Savelev10, Hu}, which poses a highly challenging numerical problem. 
However, in many cases of interest, e.g. when the junctions are sufficiently short, the in-plane stiffness of the superconducting phase suppresses spatial variations and hence enables one to treat the phase as homogeneous across the junction~\cite{Sam15, Hoeppner15, Okamoto16, Okamoto17}. 
This key simplification reduces the equations of motion to a set of ordinary differential equations. For a single junction, these are formally identical to a forced pendulum or a sliding charge density wave, both of which are well studied systems in classical chaos theory. 
Consequently, the non-equilibrium dynamics of single Josephson junctions were studied starting already in the 1970's \cite{Bohr70, MacDonald83}. The inherently strong nonlinearities give rise to period-doubling routes to chaos or strange attractors \cite{Huberman80, DHumieres82, Octavio85, Kerr85, Romeiras87, Kovacic92}. 
More recently, synchronisation of multiple Josephson junctions was studied in~\cite{Lin11} with the goal of producing coherent terahertz radiation. 

Here we investigate the dynamics of a stack of short, capacitively coupled Josephson junctions under the influence of strong driving by external laser pulses. 
In a recent paper~\cite{Schlawin17}, we investigated the switching between different macroscopic states using ultrashort terahertz pulses. In particular, we showed that the switching mechanism can be explained by the destabilisation of the uniform plasma mode. This mode can be excited by frequencies below the Josephson plasma edge, $i.e.$ the frequency at which linear plasma waves can propagate in the system. Strong driving beyond the linear regime allows nonlinear low-frequency waves to penetrate the material and affect currents inside. We further showed that this low-frequency driving can destabilise quasiparticle currents, and thus drive the system into the superconducting state. We had speculated that this might be a new surprising approach to effectively ``laser cool" the system. In this paper, we further test this hypothesis by simulating short junctions, where a long-time evolution is numerically stable. 
We analyse how the current-voltage relation can be manipulated by driving these short junctions. In contrast to earlier studies, our main focus lies on the interaction of the driving with thermally activated fluctuations.
By appropriate choice of the driving frequency and amplitude, we find that it is possible to either enhance or reduce the average voltage drop across the system in a large range of parameters. 

The paper is structured as follows: In section~\ref{sec.model}, we introduce the model and provide a brief overview of the phase dynamics of a single Josephson junction. In section~\ref{sec.results}, we then simulate a stack of Josephson junctions driven by a single-frequency external laser and analyse under which conditions the optical control of the voltage drop is feasible. Finally, we conclude in section~\ref{sec.conclusion}.

\section{Model: Stacks of short Josephson junctions}
\label{sec.model}

We consider the evolution of short, capacitively coupled Josephson junctions. Each junction is characterised by the gauge-invariant phase difference $\phi_n (\tau)$. In dimensionless units, the equations of motion are given by~\cite{Hoeppner15, Okamoto16, Okamoto17, Sam15}
\begin{align}
\frac{\partial^2 \phi_n}{\partial \tau} + \left(1 - \alpha \nabla_n^2\right) \left[ \nu_c \frac{\partial \phi_n}{\partial \tau} + \sin \phi_n\right] &= j_{\text{ext}} + \xi_n (\tau) + V_{dr} (\tau), \label{eq.toy-model}
\end{align}
where $\nu_c$ describes damping due to incoherent quasiparticle currents, $j_{\text{ext}}$ the external current flowing through the system measured in units of the maximal Josephson current, and $\alpha$ the capacitive coupling strength~\cite{Hu}. The discrete difference operator is defined as $\nabla_n^2 \phi_n = 2 \phi_n - \phi_{n+1} - \phi_{n-1}$. 
Time is measured in units of the Josephson plasma frequency $\omega_p$, $i.e.$ $\tau = \omega_p t$. Likewise, frequencies will be measured in units of the plasma frequency in the remainder of this paper. 
In the following, we set the parameters to $\alpha = 0.1$, $\nu_c = 0.1$ \cite{Savelev10, Hu}, and $j_{\text{ext}} = 0.25$, unless specified otherwise. 

In Eq.~(\ref{eq.toy-model}), we also added a driving term, which we parametrise as
\begin{align}
V_{dr} (\tau) &=  A \sin (\omega_{dr} \tau), \label{eq.cw-driving}
\end{align}
to describe phenomenologically the external diving of the Josephson phase by terahertz pulses with a driving frequency $\omega_{dr}$ and amplitude $A$. 
Similar models were recently used in~\cite{Hoeppner15, Okamoto16, Okamoto17} to describe the physics of high-temperature superconductors under the strong driving of phonons.

Additionally, we added a stochastic noise term $\xi_n (\tau)$ modelling thermal phase fluctuations. In the following, we consider white noise with $\langle \xi_n (\tau) \rangle = 0$ and
\begin{align}
\big\langle \xi_n (\tau) \xi_{n'} (\tau') \big\rangle &= X^2 \delta_{n, n'} \delta (\tau - \tau'),
\end{align}
where the intensity $X^2 \propto k_B T$ is proportional to the thermal energy in the system, $k_B T$. 
In the end, the exact temperature scale is set by the material, and in this paper we vary it between $X = 0.05$, where thermal phase slips are very rare and $X = 0.25$, where they take place very frequently. The temperature scale can then be determined by their switching distribution \cite{Sam15}. We neglect the material-dependent influence of the temperature on other system parameters, such as the quasiparticle damping $\nu_c$, as a change of their numerical values is not expected to qualitatively change the physics investigated in this paper.

In the following, we extract quantities such as the average voltage by numerically propagating Eq.~(\ref{eq.toy-model}) with a time step size $\delta \tau = 0.005$ for a total of $500,000$ steps in each run.

\subsection{Dynamics of a single undriven junction}

\begin{figure}[h]
\centering
\includegraphics[width=0.45\textwidth]{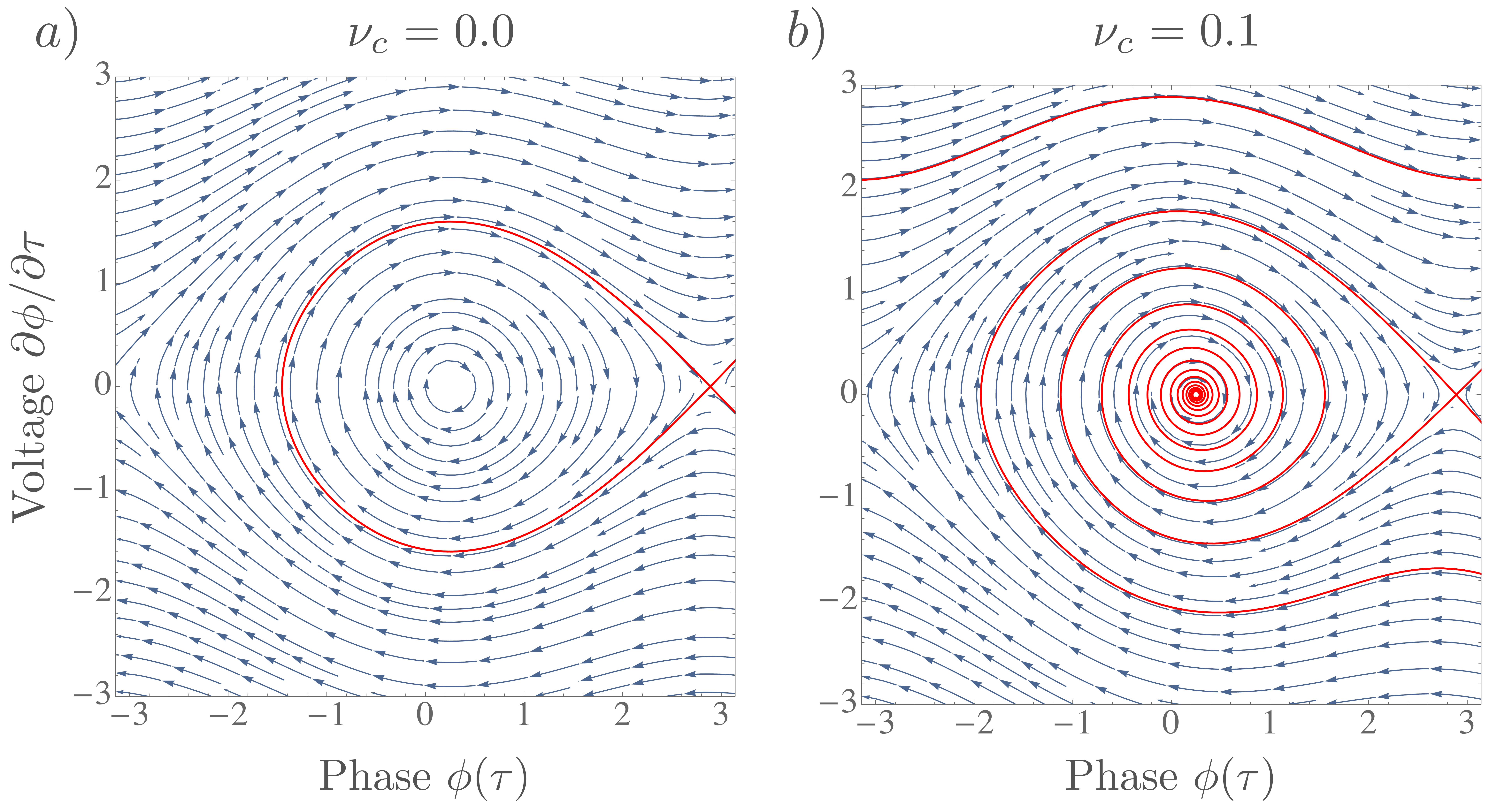}
\caption{Phase space structure of Eq. (\ref{eq.toy-model}) for a single junction with $j_{\text{ext}} = 0.25$. Blue arrows denote the local flux of the vector field, and the red lines indicate trajectories starting or ending at the unstable equilibria with a) no quasiparticle damping, $\nu_c = 0$, and b) $\nu_c = 0.1$.
}
\label{fig.phase-space}
\end{figure}

We first  consider the dynamics of a single junction. In the presence of an external current $j_{\text{ext}}$, Eq.~(\ref{eq.toy-model}) for a single junction reduces to the equation of motion of a virtual particle in a tilted washboard potential of the form  $U (\phi) = - \cos (\phi) - j_{\text{ext}} \phi$.
It has a single steady state solution,
\begin{align}
\phi_{sc} = \arcsin j_{\text{ext}}, \label{eq.phi_sc}
\end{align}
which we refer to as the superconducting state, since the voltage across the junction, $V \equiv \partial \phi / \partial \tau$, vanishes at all times. The virtual particle remains in a local minimum of the potential $U (\phi)$, and the external current is transmitted as a supercurrent.

In addition, when there is a finite damping due to quasiparticles and the current is sufficiently large, such that $j_{ext} / \nu_c \gg 1$, there also exists a stable limit cycle, the McCumber state~\cite{McCumber}. It is an approximate solution of Eq.~(\ref{eq.toy-model}),
\begin{align}
\phi_{res} (\tau) \simeq \omega_0 \tau + \Im \left\{ \frac{e^{i \omega_0 \tau}}{\omega_0^ 2 - i \omega_0 \nu_c} \right\}, \label{eq.phi_res}
\end{align} 
where $\omega_0 = j_{\text{ext}} / \nu_c$. 
Here, the energy loss due to the friction is exactly cancelled by the potential energy gain, such that the virtual particle keeps sliding down the tilted potential landscape at an approximately constant speed, given by the first term in Eq.~(\ref{eq.phi_res}). The underlying washboard potential is reflected in the oscillations described by the second term, which form a small correction to the dominant first term.
The external current is transmitted as a quasiparticle current, and the system obeys the Ohmic law $\partial \phi / \partial\tau \simeq j_{\text{ext}} / \nu_c$. Consequently, we will refer to the McCumber state~(\ref{eq.phi_res}) as the resistive state in the following. 

The phase space corresponding to a single junction is shown in Fig.~\ref{fig.phase-space}. The left panel shows the case of negligible damping, $\nu_c = 0$. We find a small area around the superconducting state~(\ref{eq.phi_sc}), where harmonic oscillations take place, and the particle remains trapped within a local minimum of the tilted washboard potential. Upon crossing the red separatrix, the trajectories become unbounded. The particle can escape the local minima and travel indefinitely. Due to the presence of the external current $j_{ext}$, the unstable equilibrium point moved away from the unstable equilibrium of the pendulum at $\phi = \pi$, just like the stable superconducting state~(\ref{eq.phi_sc}) moves away from zero. 
In the right panel, we add the damping. Consequently, the limit cycle~(\ref{eq.phi_res}) emerges at $\partial \phi / \partial\tau = 2.5$. 

\begin{figure}[t]
\centering
\includegraphics[width=0.4\textwidth]{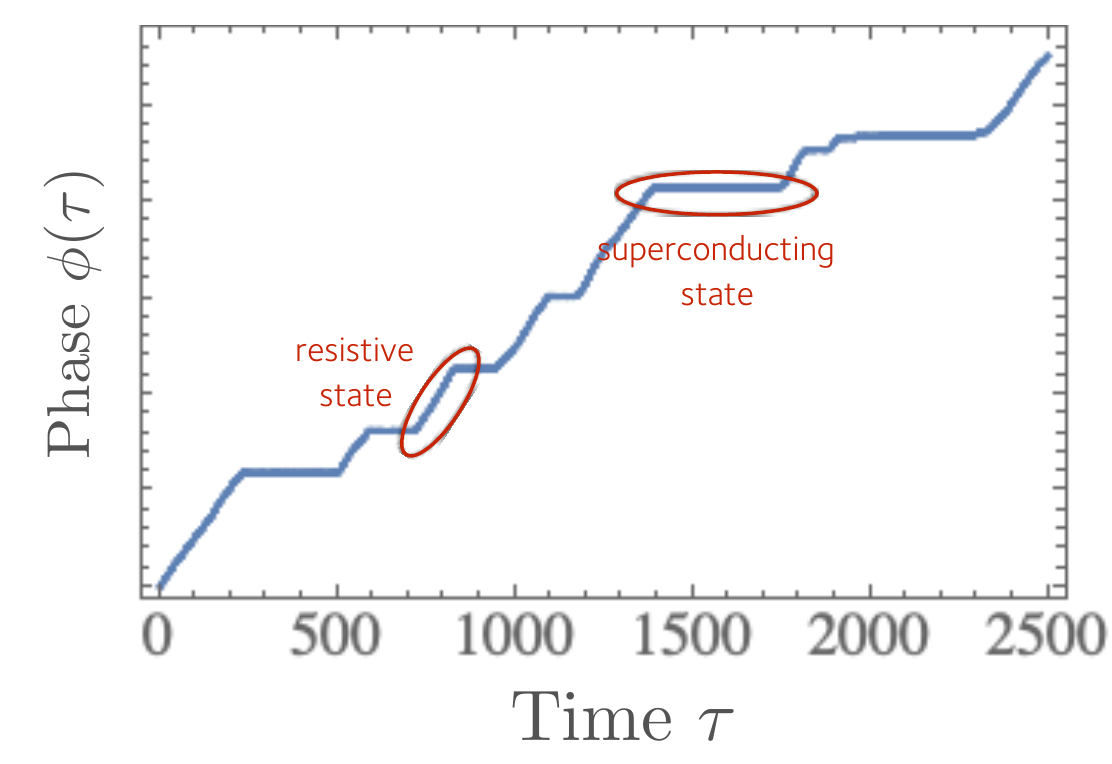}
\caption{Time evolution of a single phase according to Eq.~(\ref{eq.toy-model}). The noise is picked from a Gaussian ensemble with zero mean and standard deviation $X = 0.3$.
}
\label{fig.time-evolution}
\end{figure}

Additional noise $\xi (\tau)$ stochastically drives transitions between the two steady states. 
Fig.~\ref{fig.time-evolution} shows the time evolution of a single phase according to Eq.~(\ref{eq.toy-model}). The fluctuations (too fast to be resolved in the plot) randomly switch the system between the superconducting and the resistive state. As a consequence, an experiment at finite temperature will yield an average voltage, which is determined by the relative amount of time the system spends in each of the two steady states. Note that the fluctuation strength in Fig.~\ref{fig.time-evolution} is larger than in the rest of the paper, since the capacitive coupling between junctions strongly affects their propensity to undergo phase slips. Hence, the single junction cannot be compared directly to the stack. 

\subsection{Single driven junction}

\begin{figure}[t]
\centering
\includegraphics[width=0.49\textwidth]{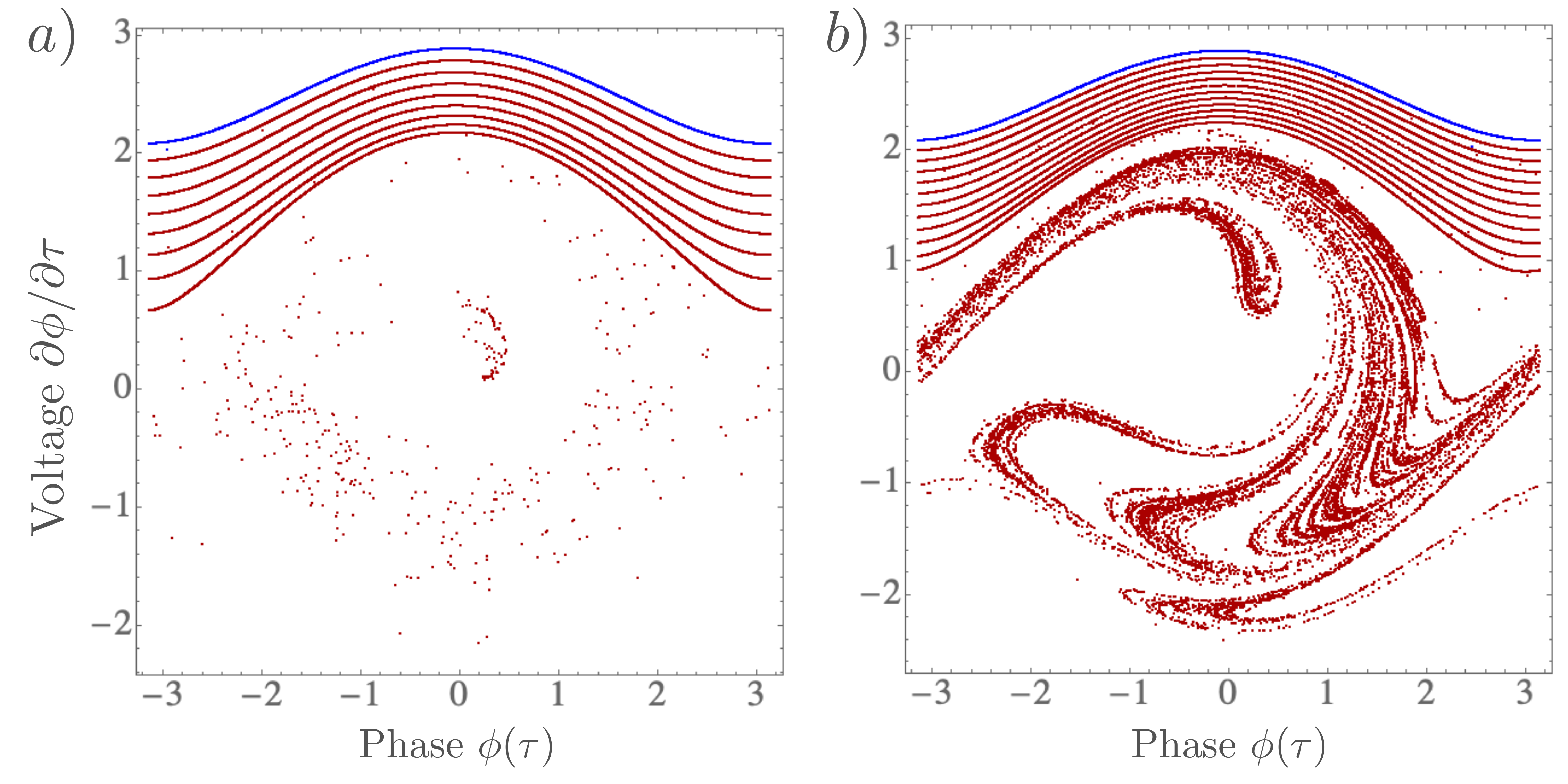}
\caption{a) Poincare section of a single junction~(\ref{eq.toy-model}) driven by Eq.~(\ref{eq.cw-driving}) with $\omega_{dr} = 0.3$.  b) The same for $\omega_{dr} = 0.5$.  In both panels, we start in the resistive state (blue dots indicate a trajectory in the undriven system), and increase the driving amplitude $A$ from 0.1 to $0.8$ in steps of $0.05$.
}
\label{fig.Poincare}
\end{figure}

We now consider the impact of periodic driving $V_{dr} (\tau)$ on the dynamics of a single junction. To simplify this discussion, we neglect the noise at first. The driving~(\ref{eq.cw-driving}) increases the dimension of the dynamical system by one. Hence, we construct Poincare sections of the trajectories to visualize the driven dynamics.
The driving introduces a new timescale $\omega_{dr}$, so we mark the state space position $(\phi, d\phi / d\tau)$ whenever $\omega_{dr} \tau = n \times 2 \pi$ for $n \in \mathds{N}$ in Fig.~\ref{fig.Poincare}. 

In Fig.~\ref{fig.Poincare}a), the blue points map the trajectory of the undriven dynamics, when starting in the resistive state. As expected, these points lie on the red limit cycle of the resistive state shown in Fig.~\ref{fig.phase-space}b). The orange dots are created by trajectories which are also initialised in the resistive state of the undriven system, but then evolve under the influence of a driving field with increasing amplitude.
Small driving amplitudes shift the trajectories along the vertical axis towards smaller values. They form trajectories which follow the undriven case, but at reduced average voltages. 
With increasing amplitude, the trajectories approach and eventually cross the separatrix in Fig.~\ref{fig.Poincare}b), which separates the unbounded, resistive trajectories from localised, superconducting ones. Consequently, in Fig.~\ref{fig.Poincare}a), the system spirals down towards the sc state, which in phase space forms an attractor at $\phi = \arcsin (j_{ext})$ (see Eq.~(\ref{eq.phi_sc})). In panel Fig.~\ref{fig.Poincare}b), when driving at a higher frequency, the average voltage is also reduced until the resistive state becomes unstable. However, the system no longer relaxes towards the sc state, but instead converges to a strange attractor. This attractor, whose shape depends on the driving frequency as well as the other system parameters, has been described elsewhere~\cite{Huberman80}, and is sometimes referred to as a phase-locked state, as the phase across the junction is synchronised with the driving field. 
For our purposes, it is important to note that the average voltage recorded in this state is determined by both the frequency and the amplitude of the driving field. 
When we increase the driving frequency, we find that the attractor is pushed away from the sc equilibrium point. Once it merges with the separatrix, the attractor disappears and only resistive dynamics remain stable.

This discussion shows that - at least in principle - it should be possible in a large parameter regime to reduce the voltage of the system in the resistive state. Conversely, when the system is initialised in the superconducting state, adequate driving may force it into the resistive or the phase-locked state.
At larger driving amplitudes, one might either destroy the resistive state and switch the system into the superconducting state or reduce the voltage by forcing the system into a synchronised state, where the voltage is set by the driving frequency. In the following, we will investigate how this control can be achieved in the presence of thermal fluctuations and capacitive coupling between the junctions.

\section{Results}
\label{sec.results}

\subsection{Current-voltage relations of the undriven system}

\begin{figure}[ht]
\centering
\includegraphics[width=0.45\textwidth]{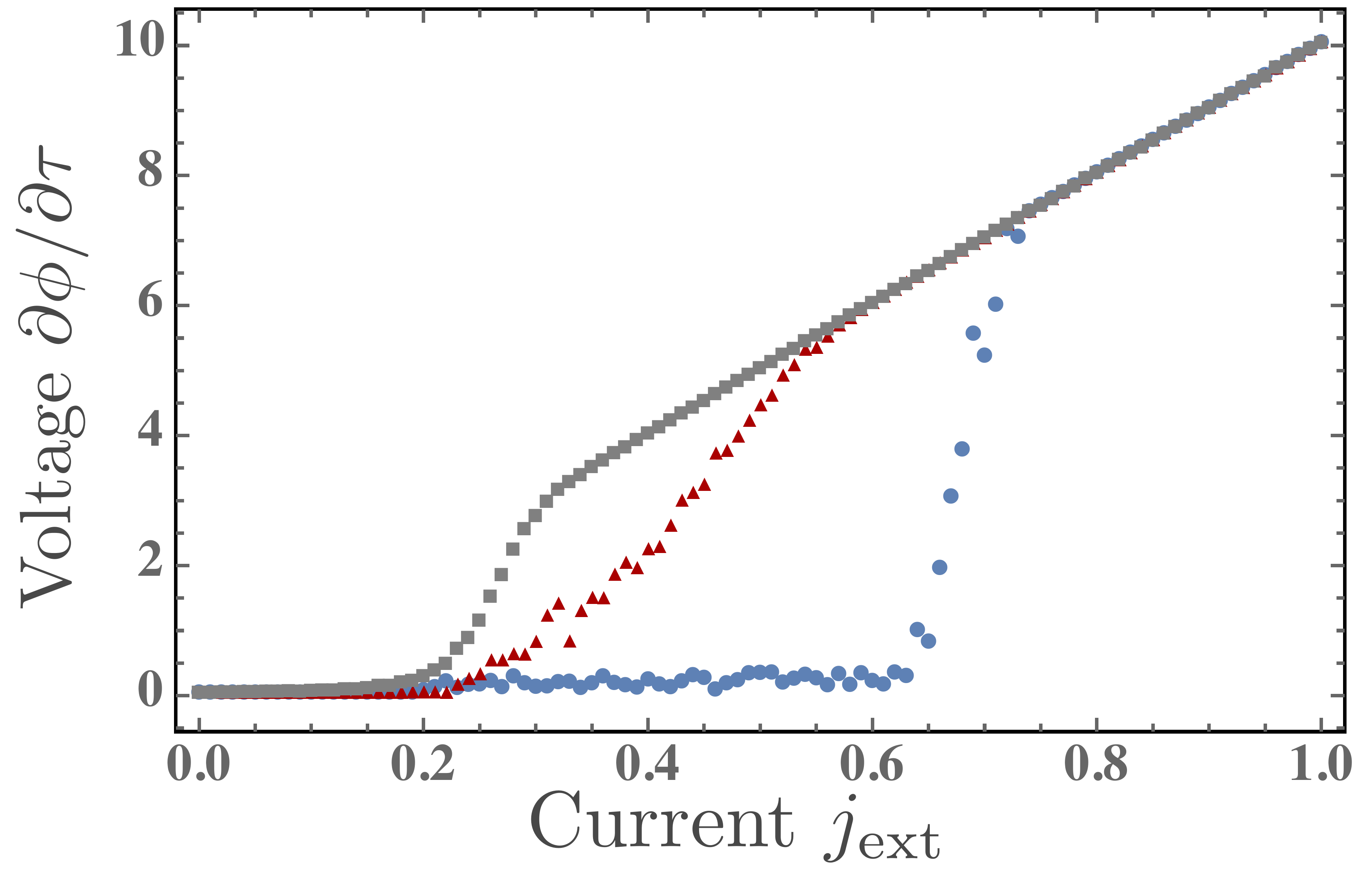}
\caption{Current-voltage relation in a stack of 100 junctions with fluctuations set to $X = 0.05$ (blue points), $0.15$ (red triangles), and $0.25$ (gray squares). We initialize the system such that each junction is chosen to be in a resistive state, if a random number is smaller than the noise amplitude $X$, and in a superconducting state otherwise. After an initial propagation period to reach a steady state, the voltage is recorded.
}
\label{fig.undriven}
\end{figure}

In Fig.~\ref{fig.undriven}, we show simulations of the current-voltage relation of a stack of capacitively coupled Josephson junctions under the influence of noise. As the current increases, the potential barrier between the superconducting state~(\ref{eq.phi_sc}) and the unstable equilibrium (see Fig.~\ref{fig.Poincare}), which thermal noise has to supply in order to switch the system from the superconducting to the resistive state, decreases. In contrast, the amount of energy needed to drive the system from the resistive state back to the superconducting state increases. As a consequence, the system will be more and more likely to find itself trapped in the resistive state.
At low noise levels (low temperatures) with $X = 0.05$, the system remains superconducting, $i.e.$ at zero voltage, until $j_{ext} \simeq 0.65$. It then quickly switches to the resistive state, where it remains for $j_{ext} \gtrsim 0.7$. At large noise levels corresponding to high temperatures with $X = 0.25$, this switch already occurs above $j_{ext} \simeq 0.2$, $i.e.$ almost as soon as the resistive state~(\ref{eq.phi_res}) becomes a stable solution of Eq.~(\ref{eq.toy-model}). Finally, at intermediate temperatures, there is an extended range of currents, $j_{ext} = 0.2 \ldots 0.5$, during which the system shows a gradual shift from the superconducting to the resistive voltage. In this regime, we observe dynamics akin to those shown in Fig.~\ref{fig.time-evolution} for a single junction, where thermal fluctuations frequently switch the system between superconducting and resistive states. As a consequence, the average voltage reaches a level somewhere in between the two extrema, and we observe large fluctuations around this mean. It is in this regime, where one would expect that even weak driving can strongly influence the dynamics, as we will investigate in the following.

\subsection{Current-voltage relations of the weakly driven system}

\begin{figure}[t]
\centering
\includegraphics[width=0.45\textwidth]{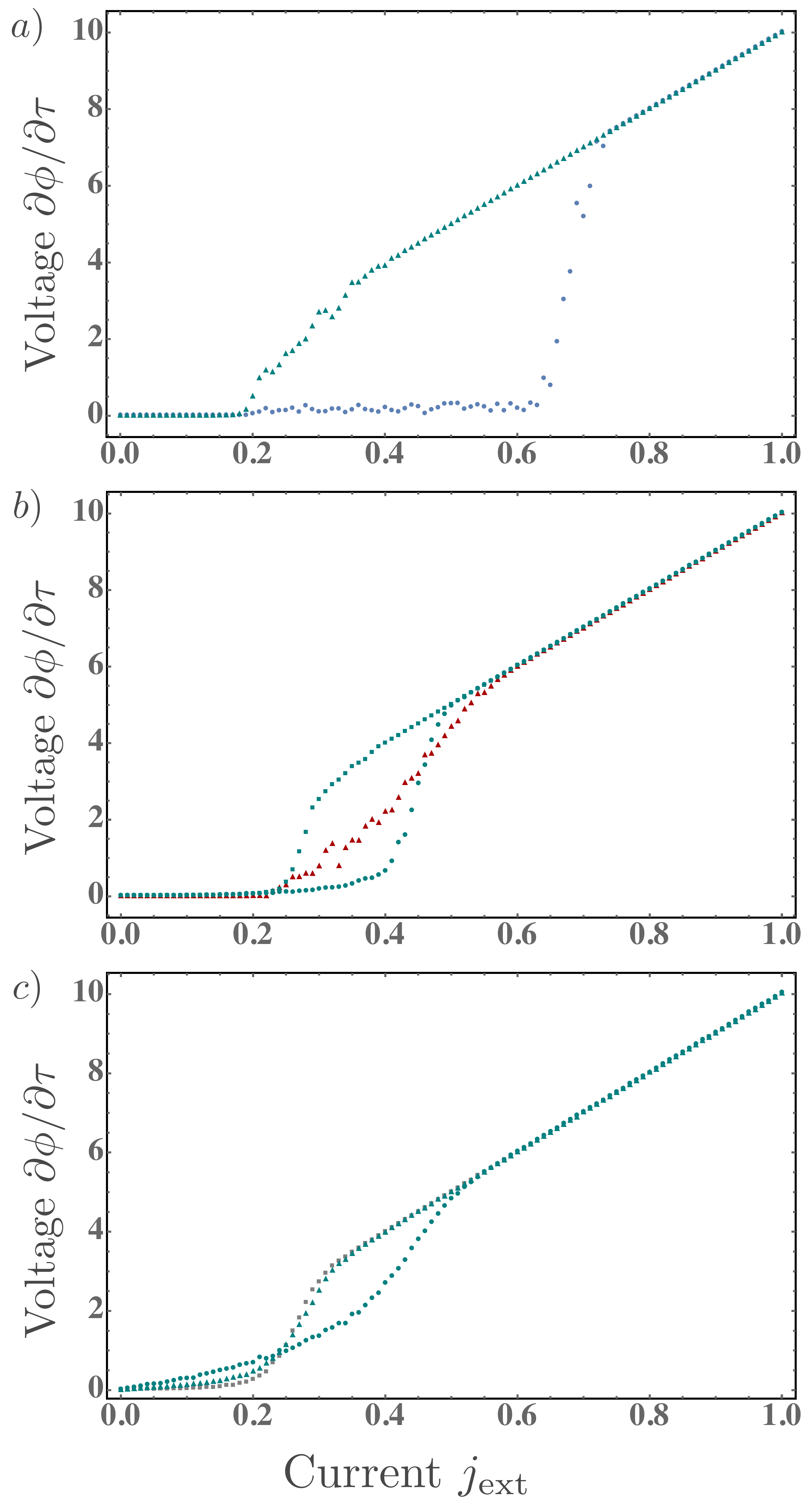}
\caption{a) Current-voltage relation in a stack of 100 junctions with weak thermal fluctuations, $X = 0.05$ (blue points), and in the presence of an external drive with amplitude $A = 0.3$  and frequency $\omega_{dr} = 0.95$ (cyan triangles).
b) the same as a), but with stronger thermal fluctuations, $X = 0.15$ (red triangles), and in the presence of an external drive with $A = 0.3$ and $\omega_{dr} = 0.05$ (cyan points) or $\omega_{dr} = 0.5$ (cyan squares).
c) the same as b), but with even stronger thermal fluctuations with $X = 0.25$.}
\label{fig.driven}
\end{figure}

\begin{figure*}[ht]
\centering
\includegraphics[width=0.95\textwidth]{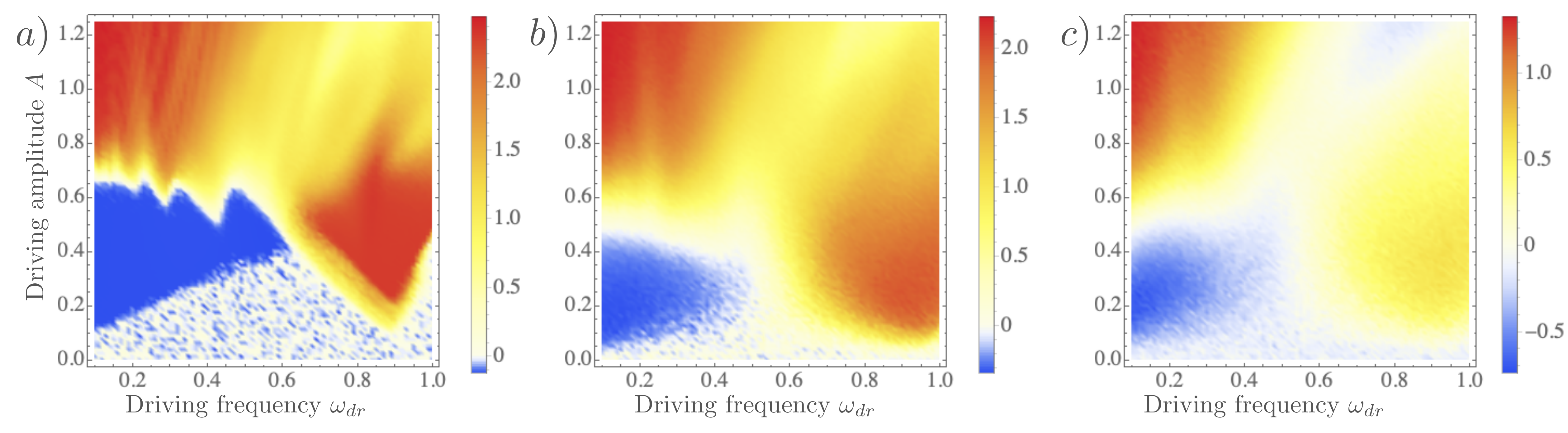}
\caption{a) Voltage difference relative to an undriven stack as a function of the driving frequency $\omega_{dr}$ and amplitude $A$ with weak thermal fluctuations, $X = 0.05$ and $j_{ext} = 0.25$.
b) the same as a), but with stronger thermal fluctuations, $X = 0.15$.
c) the same as a), but with even stronger thermal fluctuations with $X = 0.25$.
Note that the colour code is adjusted in each plot such that no changes to the undriven voltage correspond to white colour. }
\label{fig.2d-parameters}
\end{figure*}

In Fig.~\ref{fig.driven}, we investigate how driving can influence the dynamics for each of the three temperature regimes shown in Fig.~\ref{fig.undriven}.
In the low-temperature regime in Fig.~\ref{fig.driven}a), the undriven system remains in the superconducting state up to fairly large currents. Consequently, the only thing driving at moderate amplitudes can achieve in this regime is to increase the voltage by driving the system into the resistive state. This effect turns out to be most pronounced at larger frequencies close to the plasma edge. 
At intermediate temperatures shown in Fig.~\ref{fig.driven}b), it is possible to both enhance or reduce the voltage in the transition regime. Low frequency driving (circular points) reduces the average voltage between $j_{ext} \simeq 0.2 \ldots 0.5$, while higher-frequency driving tends to increase it. Driving at even higher frequencies can enhance the voltage further (not shown), stabilising the resistive state until the current drops below $j_{ext} \simeq 0.2$.
Similar physics is found to be at work at high temperatures in Fig.~\ref{fig.driven}c), where low-frequency driving can reduce the voltage at currents below $j_{ext} \simeq 0.5$. Interestingly though, at very low currents, where the undriven system tends to remain predominantly superconducting, the driving enhances the voltage. The high-frequency driving has very little impact on the voltage in this regime. 

\subsection{Parameter space at constant current $j_{ext}$}

To understand in more detail how the driving affects the system dynamics, we depict the change of the voltage relative to the undriven state as a function of both driving frequency and amplitude in Fig.~\ref{fig.2d-parameters}. The external current $j_{ext}$ is kept fixed. We again consider the three temperatures of previous plots in panels a)-c). 

At low temperatures, in Fig.~\ref{fig.2d-parameters}a), an intricate structure can be seen most clearly, as it is not strongly blurred by fluctuations. We note in passing that the form of this structure is very similar to our simulations in a previous publication~\cite{Schlawin17}, where the full spatially dependent phase difference at zero temperature was simulated in synchronised stacks of junctions. This further demonstrates the applicability of our model to the problem at hand. 
In Fig.~\ref{fig.2d-parameters}a), we find a narrow strip (the white region) at very weak driving amplitudes below $A \sim 0.1$, where the voltage is not affected by the driving. Then we find an extended region between $A \simeq 0.2 \ldots 0.6$, where at low driving frequencies $\omega_{dr} \lesssim 0.6$ the driving actually suppresses the weak thermal fluctuations and slightly reduces the resulting voltage. At frequencies $\omega_{dr} \gtrsim 0.6$, there is a sharp transition to a strong enhancement of the voltage. This region corresponds to the driven case in Fig.~\ref{fig.driven}a) that forces the system into the resistive state. Energy is pumped resonantly into the system, until the superconducting state becomes unstable~\cite{DHumieres82}.
The same voltage can be reached also with low driving frequencies $\omega_{dr} \lesssim 0.4$ and large amplitudes $A \gtrsim 0.6$. These two regions with maximal voltage are separated by a pronounced intermediate minimum stretching roughly between $\omega_{dr} = 0.5$ and $A = 0.6$ to $\omega_{dr} = 0.8$ and $A = 1.2$. This region is dominated by synchronised states (compare the parameters in Fig.~\ref{fig.Poincare}b)), and correspond to an emergent state due to the driving that cannot be found in the equilibrium material. 

At higher temperatures, this structure of the parameter space persists, even though it is blurred more strongly by thermal fluctuations, and fine features at the boundary of distinct regions are washed out. As the thermally induced voltage in the undriven system increases in panels b) and c), so does the possible reduction of said voltage. Generally, we find an extended region at low frequencies and moderate driving amplitudes that suppresses the voltage. Hence, superconducting currents can be stabilised in this parameter regime by destabilising the competing resistive channel. 
Furthermore, the region dominated by phase-locked states also remains visible at higher temperatures. Since the phase-locked voltage becomes comparable to the thermal one, these will, however, become more difficult to detect experimentally. 

\subsection{Dependence on the driving amplitude}

\begin{figure}[ht]
\centering
\includegraphics[width=0.45\textwidth]{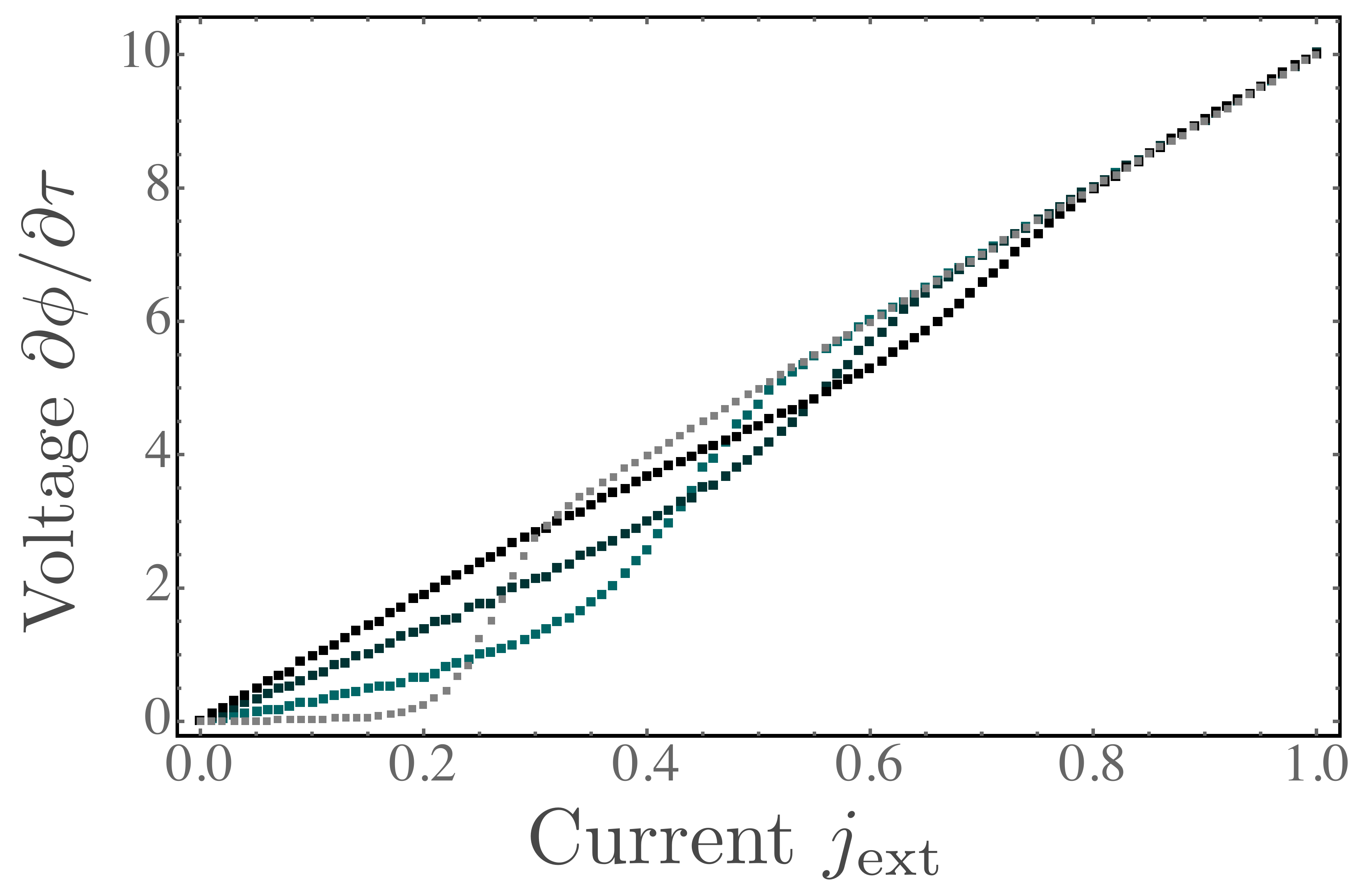}
\caption{Current-voltage relation in a stack of 100 junctions with strong thermal fluctuations, $X = 0.25$ (gray squares), and in the presence of an external drive with amplitudes $A = 0.4, 0.6,$ and $0.8$ and frequency $\omega_{dr} = 0.1$ (cyan squares with increasingly darker shade), respectively.
}
\label{fig.driven2}
\end{figure}

Finally, in Fig.~\ref{fig.driven2} we investigate how the driving field amplitude $A$ influences the current-voltage relation at high temperature. The driving frequency is fixed. 
At small amplitudes, the voltage can be reduced most strongly. However, the impact is restricted to rather small currents close to the critical current, where the undriven system switches to the resistive state naturally. To increase the maximal currents, where the voltage can be manipulated by driving, one has to increase the driving amplitude. Clearly, this comes at the expense of the maximal possible reduction. While the weakest amplitude in Fig.~\ref{fig.driven2} can achieve a maximal voltage reduction of around 50~\% for $j_{ext} \simeq 0.3$, this value drops steadily and the maximal reduction induced by the largest driving amplitude is $\sim 10$~\%. However, this reduction takes place at much larger currents, $j_{ext} \simeq 0.6$. Hence, there appears to be a clear trade-off between how strongly the voltage can be reduced and the strength of the current, at which this reduction is to be achieved - and hence the necessary driving amplitude. 

\section{Conclusions}
\label{sec.conclusion}

To conclude, we have investigated systematically the strong driving of interlayer plasma oscillations by laser fields in short stacked superconductors. In particular, we focussed on how the driving can manipulate the thermally induced voltage drop in the presence of currents. 
The key ingredient enabling this control is the strong nonlinearity of the Josephson plasmon, and its interaction with currents flowing through the system. 
At transient temperatures, we found that a substantial reduction of the thermally induced voltage drop to almost zero is possible. Thermal fluctuations can be effectively cooled by appropriate driving in this regime. At large driving amplitudes, phase-locked states can be induced, where the phase difference follows the laser pulse. This allows the reduction of the voltage to a finite but appreciatly smaller value even in the presence of very large currents. However, we found a trade-off between the amount of voltage reduction that is feasible and the strength of the currents. 

In future research, it will be interesting to investigate whether the coupling to the quantum-optical fields in cavities could yield additional control over the electronic phase of the material~\cite{ Curtis18, Sentef18, Kiffner18, Mazza19, Allocca19, Schlawin19, Kiffner19}.

\begin{acknowledgements} 
The research leading to these results has received funding from the European Research Council under the European Union's Seventh Framework Programme (FP7/2007-2013) Grant Agreement No. 319286 Q-MAC. 
\end{acknowledgements}


\begin{thebibliography}{99}


\bibitem{Kampfrath11} T. Kampfrath et al., \href{http://www.nature.com/nphoton/journal/v5/n1/full/nphoton.2010.259.html}{Nature Photon. \textbf{5}, 31--34 (2011).}

\bibitem{Kampfrath13} T. Kampfrath, K. Tanaka and K. A. Nelson, \href{http://dx.doi.org/10.1038/nphoton.2013.184}{Nature Photon. \textbf{7}, 680--690 (2013).}

\bibitem{Nicoletti16} D. Nicoletti and A. Cavalleri, \href{https://doi.org/10.1364/AOP.8.000401}{Adv. Opt. Photon. \textbf{8}, 401-464 (2016).}

\bibitem{Forst11} M. F\"{o}rst et al., \href{http://dx.doi.org/10.1038/nphys2055}{Nature Phys. \textbf{7}, 854--856 (2011).}

\bibitem{Subedi14} A. Subedi, A. Cavalleri and A. Georges, \href{http://link.aps.org/doi/10.1103/PhysRevB.89.220301}{Phys. Rev. B \textbf{89}, 220301(R) (2014).}

\bibitem{Knap15} M. Knap, M. Babadi, G. Refael, I. Martin and E. Demler \href{http://link.aps.org/doi/10.1103/PhysRevB.94.214504}{Phys. Rev. B \textbf{94}, 214504 (2016).}

\bibitem{Sentef16} M. A. Sentef, A. F. Kemper, A. Georges and C. Kollath, \href{http://link.aps.org/doi/10.1103/PhysRevB.93.144506}{Phys. Rev. B \textbf{93}, 144506 (2016).}

\bibitem{Millis17} D. A. Kennes, E. Y. Wilner, D. R. Reichman and A. J. Millis, \href{http://dx.doi.org/10.1038/nphys4024}{Nature Phys. \textbf{13}, 479 (2017).}

\bibitem{Rini07} M. Rini et al., \href{http://dx.doi.org/10.1038/nature06119}{Nature \textbf{449}, 72--74 (2007).}

\bibitem{Liu12} M. Liu et al., \href{http://dx.doi.org/10.1038/nature11231}{Nature \textbf{487}, 345--348 (2012).}

\bibitem{Nova17} T. F. Nova et al., \href{http://www.nature.com/nphys/journal/v13/n2/full/nphys3925.html}{Nature Phys. \textbf{13}, 132--136 (2017).}

\bibitem{Fausti11} D. Fausti et al., \href{http://science.sciencemag.org/content/331/6014/189}{Science \textbf{331}, 189-191 (2011).}


\bibitem{Forst14} M. F\"{o}rst et al., \href{https://link.aps.org/doi/10.1103/PhysRevLett.112.157002}{Phys. Rev. Lett. \textbf{112}, 157002 (2014).}

\bibitem{Forst14b} M. F\"{o}rst et al., \href{https://link.aps.org/doi/10.1103/PhysRevB.90.184514}{Phys. Rev. B \textbf{90}, 184514 (2014).}

\bibitem{Mankowsky17} R. Mankowsky et al., \href{https://link.aps.org/doi/10.1103/PhysRevLett.118.116402}{Phys. Rev. Lett. \textbf{118}, 116402 (2017).}

\bibitem{Hu14} W. Hu et al., \href{http://www.nature.com/nmat/journal/v13/n7/full/nmat3963.html}{Nature Mater. \textbf{13}, 705 - 711 (2014).}

\bibitem{Mankowsky14} R. Mankowsky et al., \href{http://dx.doi.org/10.1038/nature13875}{Nature \textbf{516}, 71--73 (2014).}

\bibitem{Mitrano16} M. Mitrano et al., \href{http://www.nature.com/nature/journal/v530/n7591/full/nature16522.html}{Nature \textbf{530}, 461-464 (2016).}

\bibitem{Bukov15} M. Bukov, L. D'Alessio and A. Polkovnikov, \href{https://doi.org/10.1080/00018732.2015.1055918}{Adv. Phys. \textbf{64}, 139-226 (2015).}

\bibitem{Bloch12} I. Bloch, J. Dalibard, S. Nascimb\`{e}me, \href{https://doi.org/10.1038/nphys2259}{Nature Phys. \textbf{8}, 267 (2012).}

\bibitem{Hoeppner15} R. H\"{o}ppner, B. Zhu, T. Rexin, A. Cavalleri and L. Mathey, \href{http://link.aps.org/doi/10.1103/PhysRevB.91.104507}{Phys. Rev. B \textbf{91}, 104507 (2015).}

\bibitem{Okamoto16} J.-i. Okamoto, A. Cavalleri and L. Mathey, \href{http://link.aps.org/doi/10.1103/PhysRevLett.117.227001}{Phys. Rev. Lett. \textbf{117}, 227001 (2016).}

\bibitem{Okamoto17} J.-i. Okamoto, W. Hu, A. Cavalleri and L. Mathey, \href{https://link.aps.org/doi/10.1103/PhysRevB.96.144505}{Phys. Rev. B \textbf{96}, 144505 (2017).}

\bibitem{Dienst11} A. Dienst et al., \href{http://www.nature.com/nphoton/journal/v5/n8/full/nphoton.2011.124.html}{Nature Photon. \textbf{5}, 485-488 (2011).}

\bibitem{Dienst13} A. Dienst et al., \href{http://www.nature.com/nmat/journal/v12/n6/full/nmat3580.html}{Nature Mater. \textbf{12}, 535-541 (2013).}

\bibitem{Rajasekaran15} S. Rajasekaran et al., \href{http://dx.doi.org/10.1038/nphys3819}{Nature Phys. \textbf{12}, 1012--1016 (2016).}

\bibitem{Rajasekaran18} S. Rajasekaran, J. Okamoto, L. Mathey, M. Fechner, V. Thampy, G. D. Gu and A. Cavalleri, \href{https://science.sciencemag.org/content/359/6375/575}{Science \textbf{359}, 575 (2018).}

\bibitem{Sam15} S. J. Denny, S. R. Clark, Y. Laplace, A. Cavalleri and D. Jaksch, \href{http://link.aps.org/doi/10.1103/PhysRevLett.114.137001}{Phys. Rev. Lett. \textbf{114}, 137001 (2015).}

\bibitem{Schlawin17} F. Schlawin, A. S. Dietrich, M. Kiffner, A. Cavalleri and D. Jaksch, \href{https://link.aps.org/doi/10.1103/PhysRevB.96.064526}{Phys. Rev. B \textbf{96}, 064526 (2017).}

\bibitem{Savelev10} S. Savel'ev, V. A. Yampol'skii, A. L. Rakhmanov and F. Nori, \href{http://stacks.iop.org/0034-4885/73/i=2/a=026501}{Rep. Prog. Phys. \textbf{73}, 026501 (2010).} 

\bibitem{Hu} X. Hu and S.-Z. Lin, \href{http://stacks.iop.org/0953-2048/23/i=5/a=053001}{Superconductor Science and Technology \textbf{23}, 053001 (2010).}

\bibitem{Bohr70} T. Bohr, P. Bak and H. Mogens, \href{https://link.aps.org/doi/10.1103/PhysRevA.30.1970}{Phys. Rev. A \textbf{30}, 1970 - 1981 (1984).}

\bibitem{MacDonald83} A. H. MacDonald and M. Plischke, \href{https://link.aps.org/doi/10.1103/PhysRevB.27.201}{Phys. Rev. B \textbf{27}, 201 - 211 (1983).}

\bibitem{Huberman80} B. A. Huberman, J. P. Crutchfield and N. H. Packard, \href{https://doi.org/10.1063/1.92020}{Appl. Phys. Lett. \textbf{37}, 750 - 752 (1980).}

\bibitem{DHumieres82} D. D'Humieres, M. R. Beasley, B.A. Huberman and A. Libchaber, \href{http://link.aps.org/doi/10.1103/PhysRevA.26.3483}{Phys. Rev. A \textbf{26}, 3483 -- 3496 (1982).}

\bibitem{Octavio85} M. Octavio and C. R. Nasser, \href{https://link.aps.org/doi/10.1103/PhysRevB.30.1586}{Phys. Rev. B \textbf{30}, 1586 - 1588 (1984).}

\bibitem{Kerr85} W. C. Kerr et al., \href{https://doi.org/10.1007/BF01325387}{Zeitschr. f\"{u}r Physik B \textbf{59}, 103 - 110 (2985).}

\bibitem{Romeiras87} F. J. Romeiras and E. Ott, \href{https://link.aps.org/doi/10.1103/PhysRevA.35.4404}{Phys. Rev. A \textbf{35}, 4404 - 4413 (1987).}

\bibitem{Kovacic92} G. Kovacic and S. Wiggins, \href{http://www.sciencedirect.com/science/article/pii/0167278992900922}{Physica D \textbf{57}, 185 - 225 (1992).}

\bibitem{Lin11} S.-Z. Lin, X. Hu and L. Bulaevskii, \href{https://link.aps.org/doi/10.1103/PhysRevB.84.104501}{Phys. Rev. B \textbf{84}, 104501 (2011).}

\bibitem{McCumber} D. W. McLaughlin and A. C. Scott, \href{https://link.aps.org/doi/10.1103/PhysRevA.18.1652}{Phys. Rev. A \textbf{18}, 1652 (1978).}

\bibitem{Sentef18} M. A. Sentef, M. Ruggenthaler and A. Rubio, \href{http://advances.sciencemag.org/content/4/11/eaau6969}{Science Adv. \textbf{4}, eaau6969 (2018).}

\bibitem{Curtis18} J. B. Curtis et al., \href{https://link.aps.org/doi/10.1103/PhysRevLett.122.167002}{Phys. Rev. Lett. \textbf{122}, 167002 (2019).}

\bibitem{Kiffner18} M. Kiffner, J. R. Coulthard, F. Schlawin, A. Adavan and D. Jaksch, \href{https://link.aps.org/doi/10.1103/PhysRevB.99.085116}{Phys. Rev. B \textbf{99}, 085116 (2019).}

\bibitem{Allocca19} A. A. Allocca, Z. M. Raines, J. B. Curtis and V. M. Galitski, \href{https://link.aps.org/doi/10.1103/PhysRevB.99.020504}{Phys. Rev. B \textbf{99}, 020504 (2019).}

\bibitem{Mazza19} G. Mazza and A. George, \href{https://link.aps.org/doi/10.1103/PhysRevLett.122.017401}{Phys. Rev. Lett. \textbf{122}, 017401 (2019).}

\bibitem{Schlawin19} F. Schlawin, A. Cavalleri and D. Jaksch, \href{https://link.aps.org/doi/10.1103/PhysRevLett.122.133602}{Phys. Rev. Lett. \textbf{122}, 133602 (2019).}

\bibitem{Kiffner19} M. Kiffner, J. Coulthard, F. Schlawin, A. Ardavan and D. Jaksch, \href{https://arxiv.org/abs/1905.02044}{\underline{arXiv}: 1905.02044}


\end{thebibliography}
\end{document}